\documentclass[aps,pra,twocolumn,showpacs,superscriptaddress,groupedaddress]{revtex4-1}

\usepackage{amsmath,braket,amssymb,graphicx,float,amsthm,xcolor,outlines,mathrsfs,multirow,hhline,svg}

\usepackage{dcolumn}
\usepackage[english]{babel}
\usepackage[utf8]{inputenc}
\usepackage[T1]{fontenc}
\usepackage{listings}
\usepackage[colorinlistoftodos]{todonotes} 
\usepackage[colorlinks=true, allcolors=blue]{hyperref}
\DeclareMathOperator{\nb}{\bar{{\it n}}}

\begin{document}
\title{Coherence of quantum non-Gaussian states via nonlinear absorption of quanta}
\author{Kingshuk Adhikary}
\email{kingshuk.adhikary@upol.cz}
\affiliation{Department of Optics, Palack\'{y} University, 17. listopadu 1192/12, 771 46 Olomouc, Czech Republic}
\author{Darren W. Moore}
\email{darren.moore@upol.cz}
\affiliation{Department of Optics, Palack\'{y} University, 17. listopadu 1192/12, 771 46 Olomouc, Czech Republic}
\author{Radim Filip}
\email{filip@optics.upol.cz}
\affiliation{Department of Optics, Palack\'{y} University, 17. listopadu 1192/12, 771 46 Olomouc, Czech Republic}

\begin{abstract}
The linear and phase insensitive absorption of a single quanta via coherent interactions with a saturable system, even a single ground state qubit, is sufficient to deterministically generate quantum non-Gaussian states in an oscillator, even stimulated merely by increasing thermal oscillator energy. However, the resultant states only approach Fock states and therefore do not exhibit quantum coherence. Here we overcome this limitation using a minimal step: a nonlinear phase-insensitive absorption process added to the linear one. The coherent addition of such individually passive processes allows coherence to emerge and increase in phase space without an external drive and with minimal interaction requirements. The coherence of quantum non-Gaussian states emerges because the linear and nonlinear absorption processes are not mutually passive. In the simplest case rotationally symmetric Wigner functions of the oscillator Fock states convert their many negative regions to an extremely complex asymmetric structure in sharp contrast to the rotational symmetry of those obtained by the individual interactions. We extend this case to include an unsaturable absorber (oscillator) and analyse switching between linear and nonlinear absorptions, suitable for broad classes of experiments.
\end{abstract}

\maketitle

\section{Introduction}

A core property of quantum theory is the capacity of an individual system to form coherent superpositions~\cite{baumgratz_quantifying_2014,streltsov_colloquium_2017}. While every quantum state is a superposition in some basis, some superpositions in particular contexts are exceptionally fundamental or relevant for applications. Quantum coherence undergoes intensive research in a large variety of contexts, such as open systems where only certain superpositions survive decoherence~\cite{zurek_pointer_1981,zurek_decoherence_2003} or where observable coherences are generated via external driving~\cite{allen_optical_1987,glauber_quantum_1963}, or as a resource in quantum technologies~\cite{mccormick_quantum-enhanced_2019,hu_quantum_2019} and quantum thermodynamics~\cite{campaioli_colloquium_2023}. Especially fundamental and relevant are superpositions of energy eigenstates and the ways in which they can arise. They already have diverse and experimentally demonstrated applications in quantum phase sensing~\cite{mccormick_quantum-enhanced_2019,pan_realisation_2024} and bosonic quantum error correction~\cite{fluhmann_encoding_2019,campagne-ibarcq_quantum_2020} for quantum computing and communication.

\begin{figure}
    \centering
    \includegraphics[width=\columnwidth]{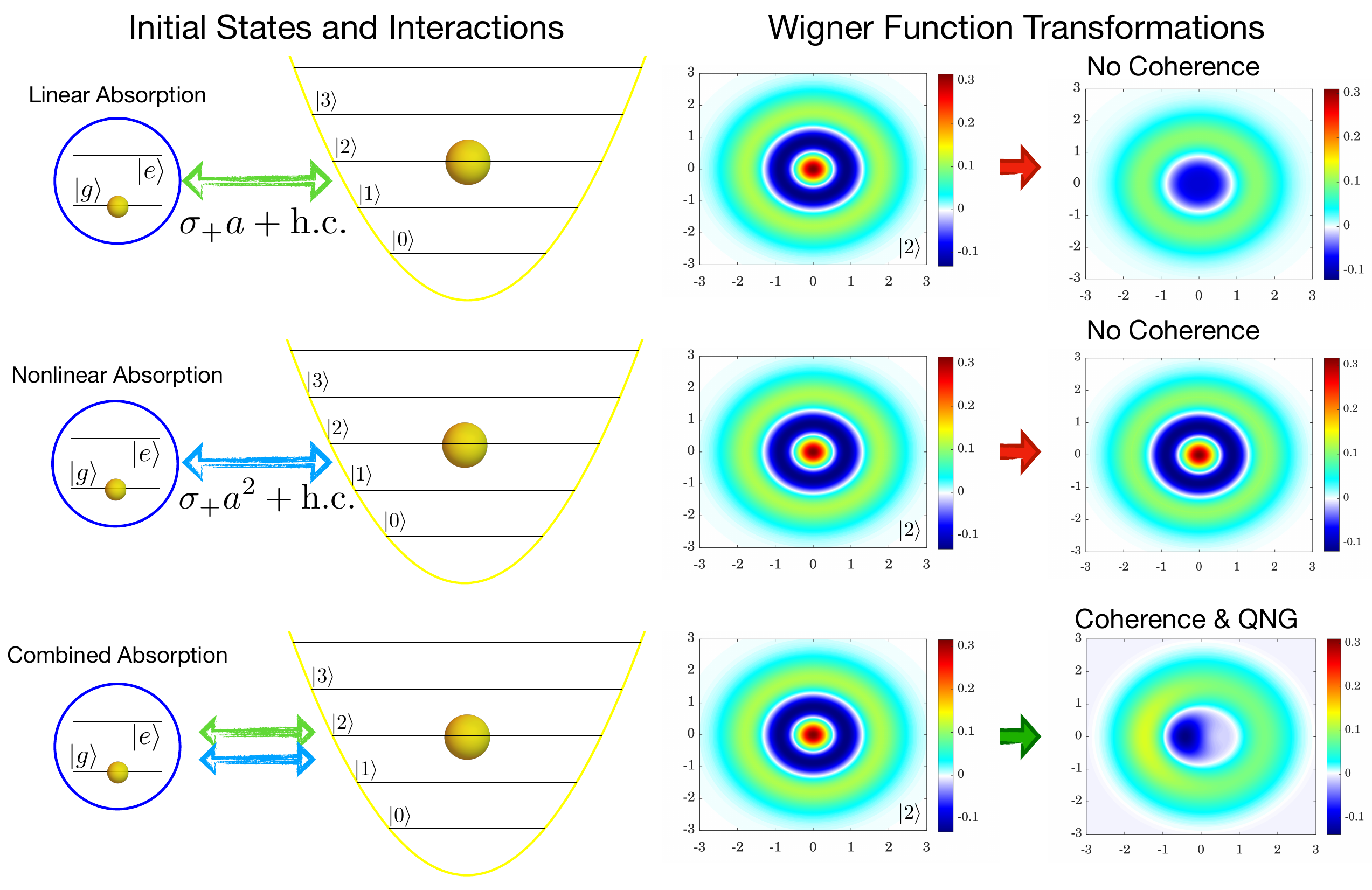}
    \caption{Coherence of quantum non-Gaussian states via a combination of linear and nonlinear absorption by a single qubit: A ground state qubit absorbing fixed quanta of energy (one or two in this illustration) from an oscillator prepared in a Fock state produces only mixtures of Fock states. These states are rotationally symmetric in phase space and therefore the coherence strictly vanishes. In contrast, if both interactions are simultaneous then a superposition of absorptions results which begins to break the rotational symmetry, indicating the emergence of quantum coherence. Wigner function transformations: The transformation of the Wigner function of Fock state $\ket{2}$ remains an incoherent mixture of Fock states for the individual interactions. The combination however results in the emergence of coherence, $\mathcal{C}=0.08$ and the loss of rotational symmetry. The Wigner function in the figure is found for interaction strengths $\frac{g^{(2)}}{g^{(1)}}=0.1$ and short interaction time $\tau=0.157$.}
    \label{Sketch}
\end{figure}

Surprisingly, basic quantum interactions deterministically generate quantum non-Gaussian oscillator states via linear coherent absorption of quanta from a thermal oscillator~\cite{marek_deterministic_2016,slodicka_deterministic_2016}. Specifically, phase insensitive and energy conserving absorption of energy from a thermal oscillator by a ground state qubit unconditionally generates states that can approach Fock states, without any external processes such as measurement, feedforward or driven/dissipative engineering. However, they cannot generate local oscillator coherence even if they do generate light-matter entanglement~\cite{filipowicz_theory_1986}. Indeed this limitation is a general property of multiphoton Jaynes-Cummings (JC) models~\cite{vogel_k-photon_1989} (see Fig.~\ref{Sketch}). 

In this paper we overcome this limitation using a nonlinear phase-insensitive absorption added to the linear one~\cite{marek_deterministic_2016,slodicka_deterministic_2016} in a fully quantum mechanical way. Thus, a combination of two different phase-insensitive absorption processes, for simplicity tested on Fock states, individually energy conserving and incapable of producing coherence, jointly results in coherent quantum non-Gaussian states. The combination is essential to create frustration between the conditions required for the interaction to be passive, i.e. energy conserving, thus allowing oscillator superpositions to deterministically develop even after ignoring the final state of the qubit. The resultant superpositions in the Fock basis show substantial quantum coherence and quantum non-Gaussian features, retaining the Wigner negativity of the original Fock states~\cite{marek_deterministic_2016,slodicka_deterministic_2016} (see Fig.~\ref{Sketch}). In what follows we demonstrate the striking extent to which this apparently simple compund interaction generates extremely quantum non-Gaussian states with substantial coherence, compare it to the classical incoherent oscillator states, extend this idea to an oscillator absorber, and suggest a feasible experiment to verify the emergence of coherence from nonlinear absorption. We close with some discussion of the nature of the nonlinear absorption interaction with respect to coherence generation.

\section{Results}

\begin{figure}
    \centering
    \includegraphics[width=\columnwidth]{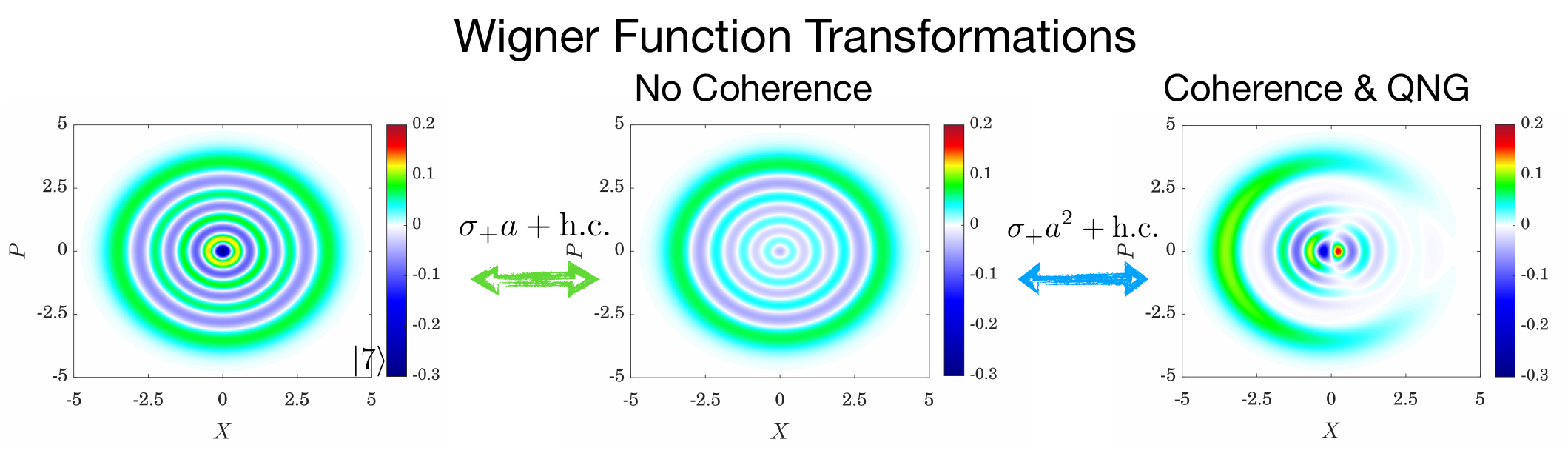}
    \caption{Sequential emergence of coherent quantum non-Gaussian states, $\mathcal{C}=0.7$ by two sequential linear and nonlinear absorptive operations. The first step, linear absorption, prepares entanglement between qubit and the oscillator but fails to produce coherence in the oscillator (see Wigner function). After the second step with a nonlinear absorption coherence already emerges and the rotational symmetry is strongly broken for higher Fock occupations. The initial state is the Fock state $\ket{7}$, the interaction time is $t=1.57$, the same for both steps, and the remaining detailed parameters are given in Fig.~\ref{BigFig}.}
    \label{switch}
\end{figure}

\begin{figure*}
    \centering
\includegraphics[width=2\columnwidth]{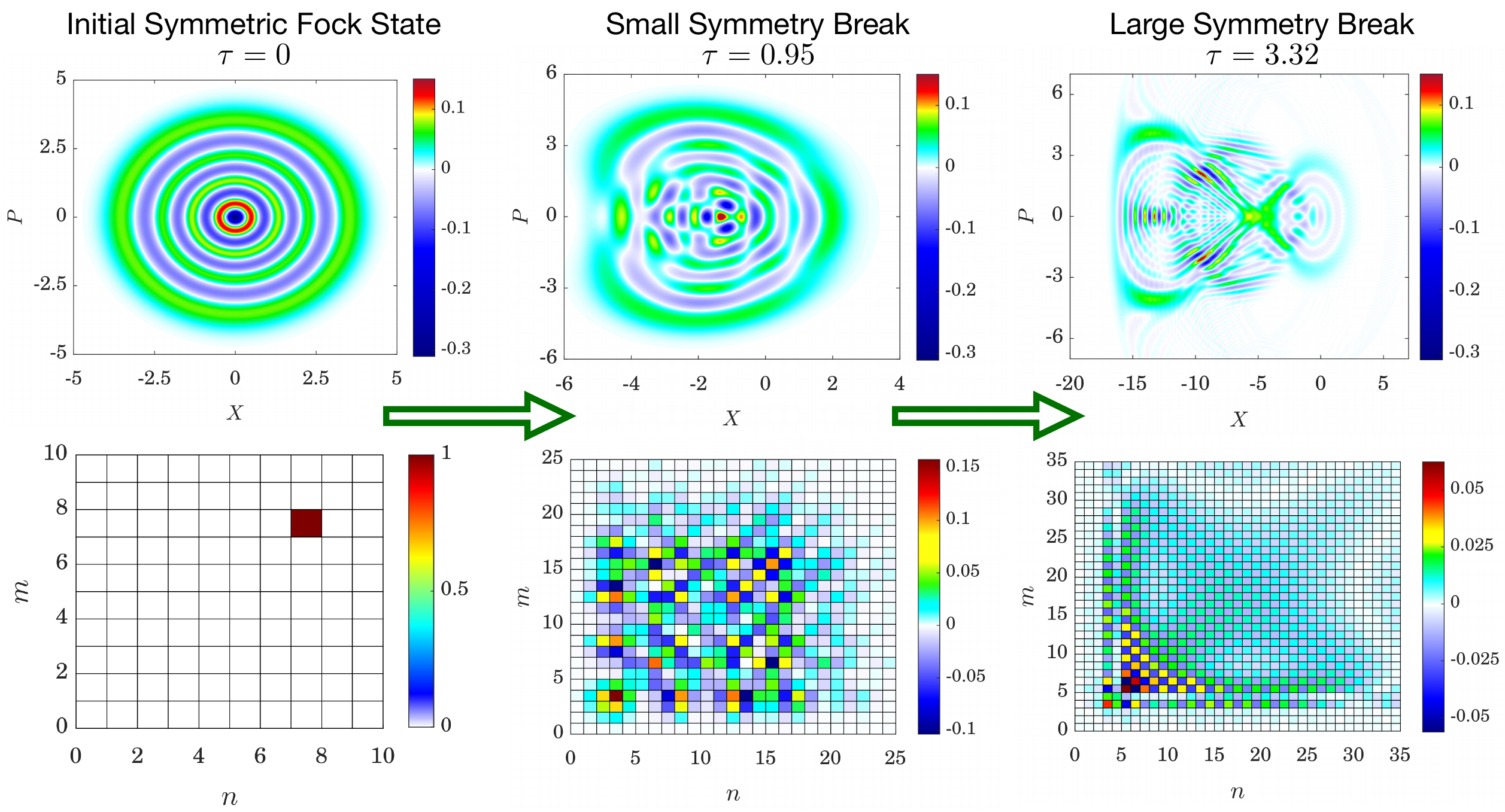}
    \caption{The emergence of coherent quantum non-Gaussian states from frequency frustrated nonlinear absorption, stimulated by initially incoherent Fock states in $b$ at a fixed coupling ratio $\frac{g^{(2)}}{g^{(1)}}=0.1$. The initial ($\tau=0$) highly non-Gaussian state is the Fock state $\ket{7}$. The rightmost state corresponds to the maximum coherence, $C\approx4$, achieved over the interval $0\le\tau\le2\pi$ at time $\tau=3.32$ and more than 4 times larger than the short time approximation discussed in the main text. The central states correspond to an example of a state with half the maximum coherence, in this case achieved at $\tau=0.95$. The states remain radically non-Gaussian, containing many negative regions and rotational symmetry is completely lost. The corresponding density matrices, with entries $\rho_{nm}$, below the Wigner functions show that $V$ tends to generate states with superpositions between large and small Fock states with entries very far from the original Fock state, and only small contributions from the ground state. More details on the parameter choices are given in the Appendix.}
    \label{BigFig}
\end{figure*}

To advance the results in \cite{marek_deterministic_2016,slodicka_deterministic_2016}, we address the emergence of local oscillator coherence via absorption of energy from the oscillator, starting from the pure incoherent Fock states approached by those methods. The relevant basis in which to examine nontrivial superpositions is therefore the energy eigenbasis of the oscillator, given by the Fock states $\ket{n}$, the eigenstates of the harmonic oscillator Hamiltonian $H_\omega=\omega b^\dagger b$, with constant frequency $\omega$. Naturally, the free evolution of the oscillator does not create superpositions from such a setup so the oscillator must interact with a new subsystem with free evolution $H_\Omega$, itself prepared in an incoherent state, diagonal in the energy eigenbasis. We will typically take this to be the ground state of $H_\Omega$, which can be approached by cooling. For a qubit subsystem we have $H_\Omega=\frac{\Omega}{2}\sigma_z$, with $\sigma_z$ a Pauli matrix.

Let us now be more precise. To quantify the overall coherence for an oscillator we use the relative entropy of coherence~\cite{zhang_quantifying_2016} defined as 
\begin{equation}
\mathcal{C}(\rho)=S(\rho_\text{diag})-S(\rho)\,,    
\end{equation}
where $S$ is the von Neumann entropy and $\rho_\text{diag}$ is the diagonal matrix containing the principle diagonal of the corresponding density matrix $\rho$. For JC-like interactions the oscillator and absorber Hamiltonians which set the energy eigenbases are $H_\omega$ (above). The $k$th order JC interaction takes the form 
\begin{equation}\label{ints}
    V^{(k)}=g^{(k)}\left(\sigma_+b^k+\sigma_-(b^\dagger)^k\right)\,,
\end{equation}
where $\sigma_\pm$ are the qubit raising and lowering operators. As said, such interactions are known to preserve the local incoherence of incoherent initial states~\cite{vogel_k-photon_1989}. Indeed the free evolution $H_0=H_\omega+H_\Omega$ commutes with $V^{(k)}$ for all $k$, provided certain frequency conditions are met. More precisely,
\begin{equation}
[H_0,V^{(k)}]=g^{(k)}(k\omega-\Omega)(\sigma_-(b^\dagger)^k-\sigma_+b^k)\,,    
\end{equation}
which is only zero for $\Omega=k\omega$. That is, the energy $N=k\sigma_+\sigma_-+a^\dagger a$ is a conserved quantity. When the interaction Hamiltonian commutes with the free Hamiltonian no local oscillator coherence emerges. We note in passing that a detuned model, with Hamiltonian $H=-\Delta a^\dagger a+\frac{\Omega}{2}\sigma_z+V^{(k)}$ and frequency $\Delta$, still does not generate local oscillator coherence. A simple method to overcome this limitation and produce oscillator coherences is to combine, in a fully quantum way, two of these energy conserving interactions [Eq.~(\ref{ints})] with different $k$.

\subsection{Short-time Emergence of Coherence and Sequential Approach}\label{HSwitch}

Let's examine the simplest case of $k=1$ and $k=2$, so that the interaction has the form $V=V^{(1)}+ V^{(2)}$. This combines linear ($k=1$) and nonlinear ($k=2$) absorption by the qubit. Apart from simplicity this is also the most relevant experimentally, as it involves the already extremely well characterised lowest order JC interactions~\cite{leibfried_quantum_2003}. For such interactions the number of excitations is no longer conserved, that is, there is frequency frustration between the competing absorption processes. Despite the analytically intractable nature of the system, some insight into the emergence of coherent quantum non-Gaussian states can be gained by considering the short time evolution of the system. An illuminating approach is to consider the first order expansion of the unitary evolution via the Baker-Campbell-Hausdorff theorem. That is, to first order in $t\ll1$ we have $U=e^{-i(V^{(1)}+V^{(2)})t}=e^{-iV^{(1)}t}e^{-iV^{(2)}t}+O(t^2)$. Terms beyond this approximation further increase the coherence. For this approximation, coherent quantum non-Gaussian states already emerge. Advantageously, this approximation also motivates Hamiltonian switching between interactions $V^{(1)}$ and $V^{(2)}$ or vice versa. This sequential method provides an alternative and immediately accessible procedure to produce coherence by combining passive and phase insensitive interactions coherently absorbing individual quanta. For example, in the context of trapped ions such interactions are generated by illuminating the ion at the $k-$th sideband. Therefore implementation of the switching protocol requires only that two such sideband are independently available and controllable in the same setup~\cite{mccormick_quantum-enhanced_2019}.

In Fig.~\ref{switch} we show the emergence of coherence at the level of the Wigner functions. Starting with the $V^{(1)}$ interaction and the state $|g\rangle |n\rangle \equiv |g, n\rangle$, $n>0$, the states $|g, n\rangle$ and $|e, n-1\rangle$ become coupled, and the typical state is a superposition of these two. Tracing out the qubit does not produce any oscillator coherence. When the Hamiltonian is switched to $V^{(2)}$, these two states decouple and couple to new states: $|g, n\rangle$ couples to $|e, n-2\rangle$, and $|e, n-1\rangle$ couples to $|g, n +1\rangle$. A typical state is now a superposition of these four, and the average total number of excitations has changed. In fact, we can explicitly write the state as
\begin{multline}\label{SwitchState}
    |\Psi_1\rangle= |g\rangle\left(\alpha(t)|n\rangle+\beta(t)|n+1\rangle\right)+\\|e\rangle\left(\gamma(t)|n-1\rangle+\delta(t)|n-2\rangle\right)
\end{multline}
where
\begin{eqnarray}
    \alpha(t)&=&\cos\left(g^{(2)}\sqrt{n(n-1)}t\right)\cos\left(g^{(1)}\sqrt{n}t\right)\\
    \beta(t)&=&-\sin\left(g^{(2)}\sqrt{n(n+1)}t\right)\sin\left(g^{(1)}\sqrt{n}t\right)\\
    \gamma(t)&=&-i\cos\left(g^{(2)}\sqrt{n(n+1)}t\right)\sin\left(g^{(1)}\sqrt{n}t\right)\\
    \delta(t)&=&-i\sin\left(g^{(2)}\sqrt{n(n-1)}t\right)\cos\left(g^{(1)}\sqrt{n}t\right)
\end{eqnarray}
When the qubit is now traced out the remaining oscillator is typically in a superposition of Fock states. Indeed each qubit eigenstate is coupled to a nonoverlapping superposition of Fock states so that the energy of the qubit no longer specifies the energy of the oscillator. Similar analyses hold for the inverted order of sequential operations, albeit with a different constraint on initial $n$. 

The Wigner function corresponding to the maximum coherence for the sequential coherence emergence is shown in Fig.~\ref{switch}. The result is a maximum qubit coherence of $\mathcal{C}\approx\ln2$ while the Wigner function loses rotational symmetry. Eq.~(\ref{SwitchState}) shows that tracing out the qubit results in a mixture of superpositions, each from a two dimensional subspace. Due to this structure the coherence is not increased by increasing $n$, in contrast with what follows in the long time emergence of coherence. Thus, the coherence of the oscillator in Eq.~(\ref{SwitchState}) is indeed bounded by the two dimensional subspace. However, coherence only mildly increases when repeating the switching procedure. Furthermore varying the individual interaction times for each step does not increase the coherence, and changing the order to start with $V^{(2)}$ decreases the ranges of times for which the maximum coherence emerges. We now move to the more autonomous dynamics without switching, where these limitations are surpassed.

\subsection{Long-time Emergence of Coherence}

Fig.~\ref{BigFig} shows typical examples of the oscillator at several stages after the compound interaction simultaneously involving $V^{(1)}$ and $V^{(2)}$. The initial state is always taken to be the incoherent state $\ket{g}\ket{n}$, where $\ket{g}$ is the ground state of the absorber and $\ket{n}$ is a Fock state of the oscillator. The ratio of coupling strengths is set to $\frac{g^{(2)}}{g^{(1)}}=0.1$, where the linear absorption still dominates, over the range $0\le\tau\le2\pi$ where $\tau=g^{(2)}t$ is a scaled time. Details on these parameter choices for the coherence dynamics are given in an Appendix. Once frustration of the energy conservation conditions is introduced via the combined linear and nonlinear absorption processes substantial coherence is gradually generated alongside strikingly complex Wigner distributions with strongly broken rotational symmetry and large density matrix coherences (off-diagonal elements of the density matrices). As the Wigner functions develop, their prominent negative regions persist despite the mixedness introduced by the tracing out of the resonant absorber. That is, the states produced do not belong to the class of states defined by the convex mixture of Gaussian states. States from this class may be non-Gaussian~\cite{genoni_quantifying_2010}, but they do not possess any Wigner negativity. Importantly, the rotational symmetry is gradually broken in time, as visible in Fig.~\ref{BigFig}, and the Wigner function approaches a completely new topology in phase space, going even beyond the complexity of those currently measured in nonlinear potentials~\cite{marti_quantum_2023}. The breaking of the rotational symmetry combines classical coherent displacement in phase space with quantum non-Gaussian symmetry breaking of the negative parts of interference effects in the Wigner function. Such complex symmetry broken structures appear close to the maximum of the mean number of quanta along with a large increase in the noise.
\begin{figure}
    \centering
    \textbf{Energy rise via composite absorption}
    \includegraphics[width=\columnwidth]{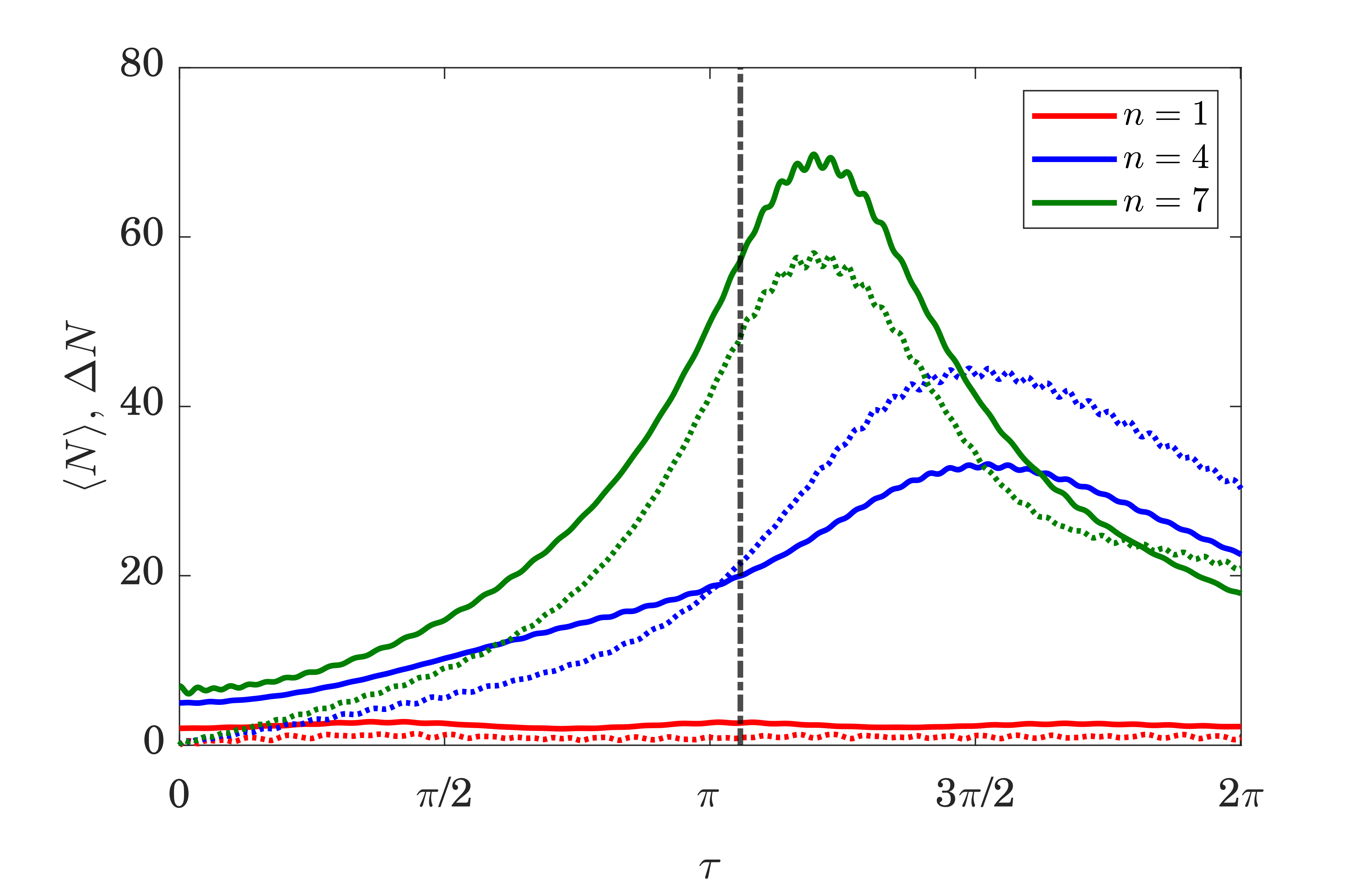}\\
    \textbf{Coherence rise via composite absorption}
    \includegraphics[width=\columnwidth]{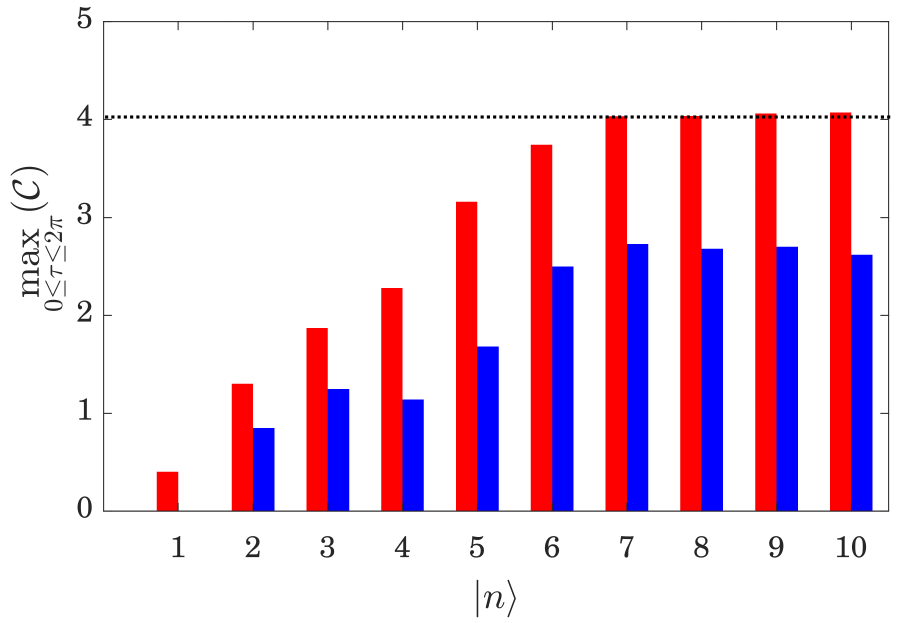}
    \caption{The spread into the Fock basis and frustration of energy conservation is captured by the rise in mean energy of the system $\braket{N}$, accompanied by a large increase in the standard deviation $\Delta N$. The maximum coherence occurs at the dashed vertical line. The bar chart shows the maximum coherence generated with $\frac{g^{(2)}}{g^{(1)}}=0.1$ over the range $0\le\tau\le2\pi$ as a function of the initial Fock state $\ket{n}$. The maximum coherence generally increases with $n$ up to saturation at $n=7$. The blue bars indicate the removal of the Gaussian shell via displacement and squeezing operations (detailed in text). The coherence persists and is thus well beyond the covariance matrix approximation.}
    \label{MeanEnergy}
\end{figure}

The value of the coherence gives only an overall view of the Fock state superpositions contributing to the coherence. Examining the density matrix coherences in Fig.~\ref{BigFig}, they spread deeply into the Fock basis, coupling low and high Fock states. This feature is not captured by the short-time approximation, nor when extended to a sequential scheme where the short-time approximation operators are repeatedly applied. Additionally, there is only a marginal contribution from the ground state. The mean number of quanta produced in the full dynamics (see Fig.~\ref{MeanEnergy}) is substantially higher than that of sequential method and the spread into the Fock basis far beyond the initial occupation number is reflected in the growth of the mean energy of the system $\braket{N}\gg7$. This effect is already known for linear absorption~\cite{podhora_unconditional_2020} but here is accompanied by the emergence of coherence. That is, linear absorption can result in an increase in mean energy, if the linear absorption is associated with blue-detuned interactions. However, it also results in a reduction of the noise in energy, as the output states closely approximate Fock states. The addition of nonlinear absorption results in a simultaneous increase in both mean energy $\braket{N}$ and noise $\Delta N$, which allows for the emergence of coherence.

Fig.~\ref{MeanEnergy} also shows the increase in maximum coherence achieved over the range $0\le\tau\le2\pi$ as a function of initial Fock state. There is a notable increase in the maximum achievable coherence with increasing $n$, up to saturation at $n=7$. The blue bars show the coherence after the removal of the Gaussian approximation, i.e. displacement and squeezing are applied until the mean values of $X=\frac{1}{\sqrt{2}}(b+b^\dagger)$ and $P=\frac{i}{\sqrt{2}}(b^\dagger-b)$ are zero and the covariance matrix is diagonal with equal entries. Quantum coherence due to Gaussian displacement/squeezing is thus removed indicating that the coherence beyond the Gaussian approximation is substantial. A negative Wigner function remains negative under Gaussian operations thus the coherence is strongly connected to the quantum non-Gaussianity of the state.

\section{Discussion}

\subsection{Weak Coupling Regimes, Dephasing and Classical Initial States}

To extend this result to many possible experimental scenarios, we analyse two significant potential obstacles in even well-isolated oscillators: the presence of free evolution alongside the compound interaction and external dephasing processes. Thus far, for simplicity, these discussions have taken place in the ultra-strong coupling regime, in which the free motion can be neglected. Reintroducing the free motion adds significant complexity to an already intractable problem. However we are interested in the emergence of coherence, rather than its optimisation. Therefore we compare the maximum coherence when the free motion is relevant with the maximum coherence obtained from our example in Fig.~\ref{BigFig}, keeping the remaining parameters unchanged. Below the saturation observed at $n=7$ it is possible to find regions outside the ultrastrong coupling regime where the coherence can be enhanced. This reflects a similar finding for qubit coherence in the Rabi model emerging from a similar unitary setting~\cite{laha_quantum_2023}. That is, this remarkable emergence of coherence is not restricted to the experimentally challenging ultrastrong coupling regime, but is much more common and may even be greater outside it. Above $n=7$ it is typical for the maximum coherence to be in the ultrastrong coupling regime. However even when the coherence is lowered the free motion does not significantly impact the quantum non-Gaussianity or complexity and negativity of the resulting Wigner functions (examples in Appendix). Similarly, the density matrices still show a substantial spread into the Fock basis coherences. 

As expected, coupling to a dephasing environment strongly reduces the coherence. We give an example in the Appendix where quantum non-Gaussian features confirmed by negativity of the Wigner function appear at shorter times but are eventually suppressed by the decoherence process, despite leaving a non-Gaussian state with nontrivial coherence. One may wonder if the emergence of coherence from nonlinear absorption is due to the nonclassical features of the Fock states or of the saturability of the qubit absorber, both of which we have relied on throughout. In fact the qualitative features of our results hold for initial states which are classical mixed states showing only thermal noise in the excitation number, as well as when the qubit is replaced by an unsaturable oscillator (see Appendix). Strikingly, the negative features of the Wigner function are more robust to initial thermal noise than to decoherence. 

\subsection{Extension of Frustration of Energy Conservation to Other Cases}

To create coherence in a single oscillator from an incoherent state the oscillator energy must change and the interaction Hamiltonian must not commute with the oscillator free motion, $[H_\Omega,V]\ne0$. With a total Hamiltonian $H=H_\omega+H_\Omega+V$, there are two distinct possibilities. Either (i) $[H_\omega+H_\Omega,V]=0$ or (ii) $[H_\omega+H_\Omega,V]\ne0$. For the first case, it follows that $[H_\Omega,V]=-[H_\omega,V]\ne0$. In this case sum of the energies of the subsystems is conserved, so that overall $V$ describes a globally passive process. In this case, even though the local excitation number of the oscillator can change, no coherence emerges. Since the total excitation number is conserved, any change in the energy of subsystem $H_\omega$ is directly compensated for by gain or loss of energy in subsystem $H_\Omega$. That is, energy exchange between the subsystems can be mediated passively by the interaction $V$, without any net exchange of energy stored in the interaction, so that there is no uncertainty in the oscillator energy. This explains why phase insensitive interactions such as JC, beamsplitters, or even trilinear interactions do not produce oscillator coherence. Passive interactions do not produce coherence in the oscillator and this holds even when the interaction is locally active. 

For the second case there are two subcases: (a) $[H_\Omega,V]=0$ and (b) $0\ne[H_\Omega,V]\ne-[H_\omega,V]$. For case (a) it is still possible to generate coherence. For example, the optomechanical interaction $a^\dagger a(b+b^\dagger)$~\cite{aspelmeyer_cavity_2014} will generate coherence in the mechanical mode $b$ even though the optical mode's free Hamiltonian commutes with the interaction and similarly for the dispersive Rabi interaction $\sigma_z(b+b^\dagger)$ in superconducting circuits~\cite{pedernales_quantum_2015,ma_quantum_2021,eickbusch_fast_2022,katz_programmable_2023} and spin-mechanics~\cite{lee_topical_2017}. This occurs even if the optical or qubit systems are not prepared in the ground state. Since their energy remains constant, yet the oscillator mode gains or loses energy, there must be an active contribution from the interaction. For the two cases above, it comes because of the counterrotating terms in the interaction; however, for our case here, we combine interactions which are each separately in the rotating wave approximation. This becomes even more appealing in case (b), which contains the nonlinear absorption interaction studied in this manuscript; moreover as with the optomechanical and dispersive Rabi interactions the nonlinear absorption method does not need to depend on the saturability of the absorber (see Appendix), as in our simple example of Eq.~\ref{ints}. Additionally our interaction is not limited to the simplest case we selected involving $k=1$ and $k=2$. Any pair of $k$ will continue to be contained in case (b), and produce coherent quantum non-Gaussian states. The frustration of the conditions for commutation to hold prevents both subsystems from conserving their energy in a nontrivial way: allowing only {\it mutually active} transformations that may result in oscillator coherence.

\section{Conclusion}

Clearly, non-Gaussian quantum superpositions in oscillators require certain minimal conditions to be met in order to arise without initial coherence, direct external coherent driving of the subsystems and counterrotating terms in the interaction Hamiltonians. Here we have used experimentally feasible phase-insensitive interactions to demonstrate some of these required conditions. That is, for this minimal case, with all subsystems prepared in incoherent states, the resulting evolution must involve a mutually active transformation in order for quantum coherence to emerge. This is not a sufficient condition, but a necessary one, so any particular Hamiltonian used like this must also be thoroughly explored for such effects.

The necessity of a mutually active transformation implies, through conservation of energy, that another physical system is at least effectively present and donating or receiving energy from the oscillator (see Appendix for discussion). In many recent cases of quantum technology, this extra physical system is in fact the external drive present during state generation that we have avoided throughout the discussion. Crucially, there are many systems in which an effective Hamiltonian dynamics can be derived which allows this source/sink of energy to be fully externalised. The effective Hamiltonian obscures the origin of quantum coherence which may require the completion of the full Hamiltonian as outlined in the appendix. Instead of searching for systems which can be externalised in this way, one may look to naturally occurring forces or technological arrangements of matter whose internal structure contains the required energy source/sink to generate coherence without external drives or counterrotating terms during the state generation. Such forces or technology may then provide a minimal approach to reservoir engineering, in which the internal structure replaces the externally driven engineered environment. This alternative transient method is a new starting point in reservoir engineering methods. Again, and in contrast to them, this method does not contain either counter-rotating terms or external coherent drives during the state preparation applied to incoherent states~\cite{poyatos_quantum_1996}. Continuing from the starting point, such minimal mechanisms can be extended for the generation of quantum non-Gaussian states without the above mentioned tools typically used in superconducting circuits~\cite{albert_holonomic_2016,ma_quantum_2021}, trapped ions~\cite{kienzler_quantum_2015}, optomechanical systems~\cite{brunelli_unconditional_2018,brunelli_linear_2019}, and two and multi-mode non-Gaussian entanglement clearly distinguishable from previously analysed cases~\cite{mcconnell_unconditional_2024}. Searching for such possibilities beyond spin-mechanical interactions with counterrotating terms~\cite{lee_topical_2017,gu_probing_2023}, and using modern technology with quantum systems may open many exciting doors in various quantum technologies requiring coherent quantum non-Gaussian states~\cite{hillery_coherence_2016,camati_coherence_2019,bernardo_unraveling_2020,ahnefeld_coherence_2022,williamson_extracting_2024}.

The high-quality Fock states used to initiate these effects are routinely available for trapped ions and superconducting circuits~\cite{podhora_unconditional_2020,podhora_quantum_2022}. However, they can also be obtained with a purely linear coherent absorption process within the rotating wave approximation~\cite{marek_deterministic_2016,slodicka_deterministic_2016}, and are thus available for testing these minimal conditions. These processes tend to result in the required Fock state in an admixture with the ground state. Even for very high contamination this does not prevent the emergence of coherence or the quantum non-Gaussian features we have discussed (see Appendix). Moreover we are not limited to such Fock states, or imperfect versions thereof, as a direct observation of these effects can also emerges from initially incoherent thermal or Poissonian oscillator statistics (see Appendix) which are also readily prepared in systems such as trapped ions and superconducting circuits.

\section{Acknowledgements}

The authors acknowledge funding from Project No. 21-13265X of the Czech
Science Foundation. This work has
received funding from the European Union’s 2020 research and innovation
programme (CSA–Coordination and Support Action, H2020-WIDESPREAD-2020-5)
under Grant
Agreement No. 951737 (NONGAUSS).

\section{Data Availability}

The data that support the findings of this study are openly available at the following URL/DOI: https:// zenodo.org/records/15392092~\cite{zenodo}.

\appendix

\section{Sequential Method}

\begin{figure}
    \centering
    \includegraphics[width=\columnwidth]{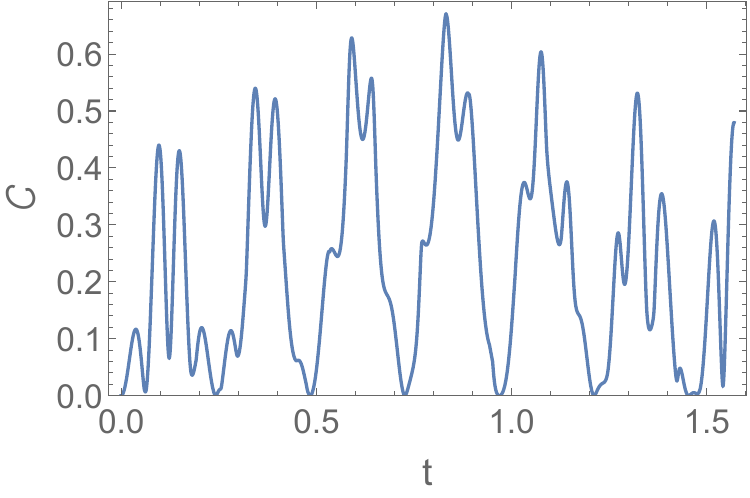}
    \caption{An example of the Hamiltonian switching process using $V^{(2)}$ to initiate the qubit-oscillator entanglement. Each interaction has the same time interval $t$, the initial oscillator occupation is $n=7$ and the ratio of coupling strengths is again $\frac{g^{(2)}}{g^{(1)}}=0.1$.}
    \label{Hswitch}
\end{figure}

We expand on the sequential scheme proposed in Section~\ref{HSwitch}. The short time approximation using the first terms of the BCH theorem is a first step which already produces coherence and breaks total excitation number conservation. First, we note the quantum non-Gaussian states that emerge from this dynamics and compare with the main result in Fig.~\ref{BigFig}. In Fig.~\ref{Hswitch} we show the effect of this Hamiltonian switching with $V^{(2)}$ initiating, $n=7$, and $\frac{g^{(2)}}{g^{(1)}}=0.1$. Repeated switching does not substantially increase the available coherence, nor does increasing the initial Fock state occupation.

\section{Coherence Dynamics}

\begin{figure*}
    \centering
    \includegraphics[width=2\columnwidth]{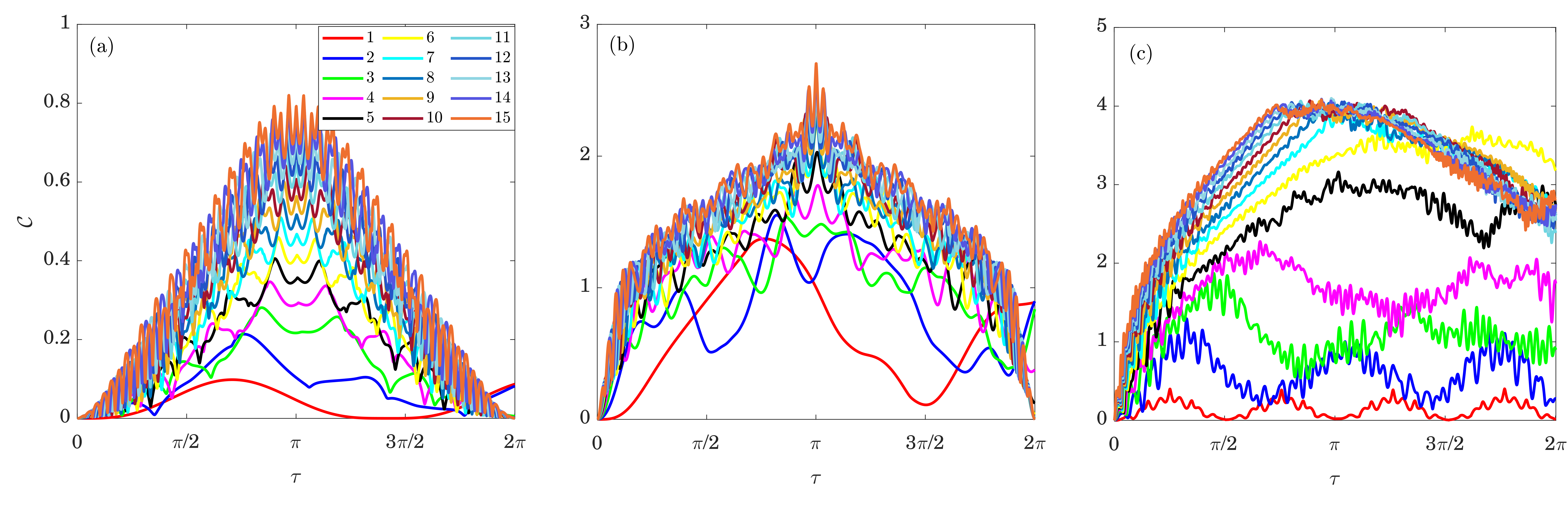}
    \caption{Coherence as a function of time for several coupling strength ratios and initial Fock states. From left to right, $\frac{g^{(2)}}{g^{(1)}}=10,~1,~0.1$. For relatively large $g^{(2)}$ the coherence dynamics has an oscillatory character, which is lost for lower $g^{(2)}$ but with a large gain in achievable coherence. The saturation with increasing $n$ is already visible here.}
    \label{CohTime}
\end{figure*}

\begin{figure}
    \centering
    \includegraphics[width=0.49\columnwidth]{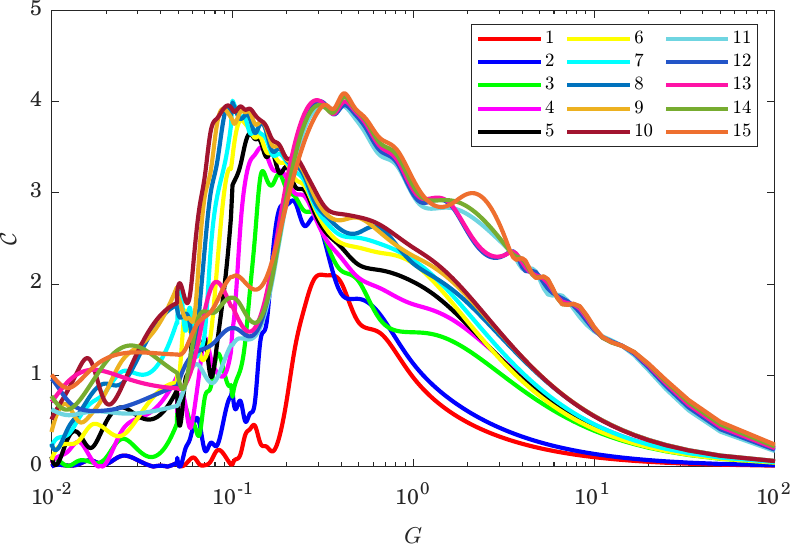}
    \includegraphics[width=0.49\columnwidth]{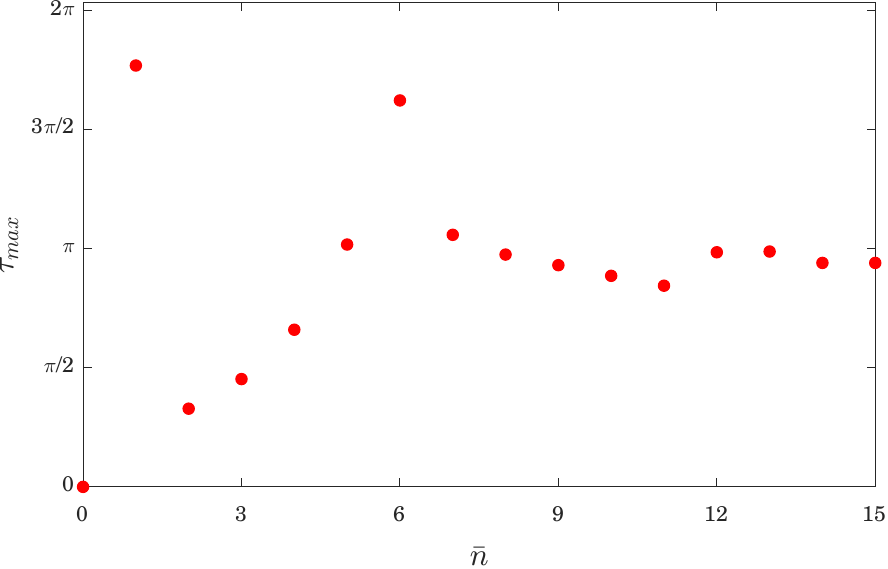}
    \caption{Top: The maximum coherence as a function of the coupling ratio $G=\frac{g^{(2)}}{g^{(1)}}$ at interaction time $\tau=\pi$. As the initial Fock occupation $n$ increases the coupling ratio at which the maximum coherence occurs converges to $G=0.1$ up to $n=10$. For higher $n$, fluctations take the $G$ away from this value, but the maximum does not change much as can be seen in Fig~\ref{CohTime}, right panel. Bottom: The interaction times $\tau$ at which the maximum coherence is extracted for $G=0.1$. These generally occur around $\tau=\pi$, although some fluctuations occur, particularly at $n=1$. However these are small, as can again be seen in Fig~\ref{CohTime}.}
    \label{CoherenceG}
\end{figure}

The coherence for the nonlinear absorption has a complex time evolution and dependence on the initial state and coupling strengths, displayed in  Fig.~\ref{CohTime}. In the main text we have selected $\frac{g^{(2)}}{g^{(1)}}=0.1$ as it produces very large coherence. The oscillatory behaviour is lost, but the magnitude of the coherence is greatly enhanced by this choice. The maximum coherence generally appears at $\tau=\pi$. When this interaction time is fixed, it becomes clear that the maximum coherence tends to occur around $\frac{g^{(2)}}{g^{(1)}}=0.1$ as $n$ increases, as in Fig.~\ref{CoherenceG}. Although some higher values of $n$ deviate from this, the increase in coherence compared to this choice of coupling strengths is negligible, as can be seen by comparing Fig.~\ref{CoherenceG} and Fig.~\ref{CohTime}. After fixing this choice, the maximum values of coherence used in the main text are those optimised over the time interval $0\le\tau\le2\pi$. Fig.~\ref{CoherenceG} also shows the times at which these maximum coherences occur as a function of the initial Fock occupation $n$. Although there are fluctuations due to minor changes in maximum coherence, this confirms the intuition that the maximum occurs around $\tau=\pi$.

\section{Weak Coupling and Decoherence}

\begin{figure*}
    \centering
    \includegraphics[width=2\columnwidth]{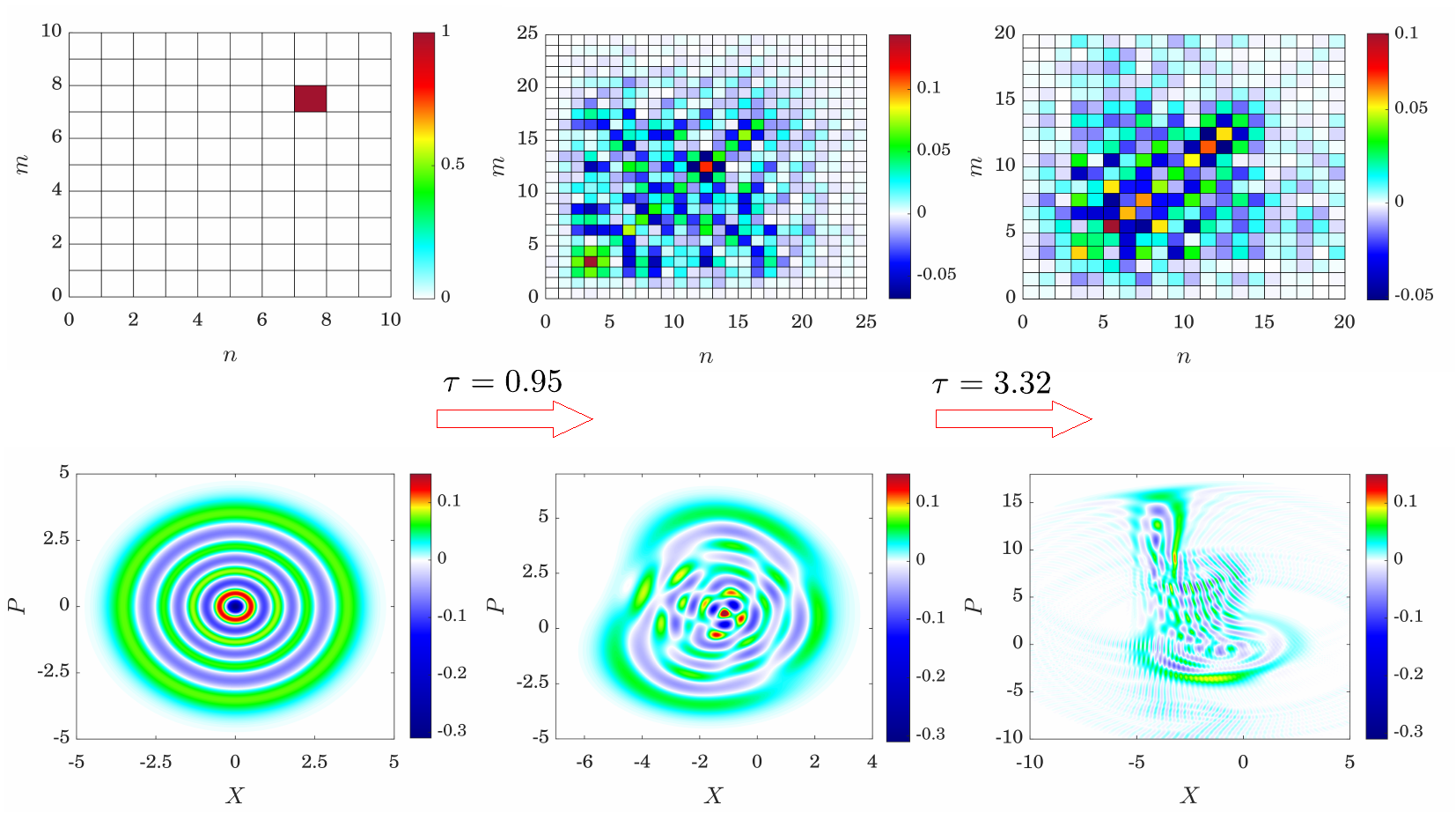}\\
    \caption{The Wigner function and density matrix for the state produced with the same parameters as the example in Fig.~\ref{BigFig}, with frequencies $\omega=\Omega=1$.}
    \label{FreeMotWigDen}
\end{figure*}

Here we give greater details on the points made in the first subsection from the discussion. Firstly, outside the ultrastrong coupling regime the effects of the free motion terms of the Hamiltonian are relevant to the coherence dynamics. As discussed in the main text the effects we have described are not limited to regimes with such large interaction strengths. Fig.~\ref{FreeMotWigDen} shows the Wigner function and density matrix of the maximum coherence for $n=7$ and $\frac{g^{(2)}}{g^{(1)}}=0.1$ with $\omega=\Omega=1$, resulting in $\mathcal{C}=3.5$. This well approximates the maximum coherence achieved in the ultrastrong coupling regime but produces quite different output states. Nevertheless the qualitative features remain: large coherence, superpositions between distant Fock states, and complex Wigner functions with multiple negative regions. 

\begin{figure*}
    \centering
    \includegraphics[width=2\columnwidth]{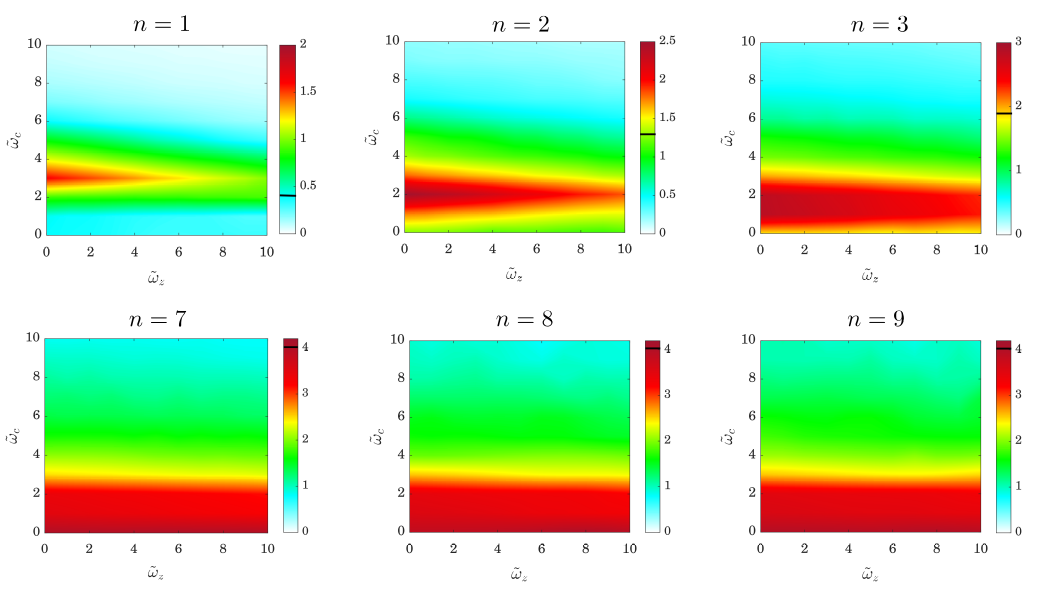}
    \caption{The maximum coherence in the free motion regime, using the same parameters as in Fig~\ref{BigFig}, with varying $n$. The black lines in the colourbars indicate the coherence extracted from the ultrastrong coupling regime. Before saturation at $n=7$, it is typical for this coherence to be much lower than that achieved in the weak coupling regime. After saturation, this behaviour changes the the maximum is close to the ultrastrong coupling regime.}
    \label{CohContour}
\end{figure*}

Notably, the large values of coherence do not simply occur close to the ultrastrong coupling regime, as demonstrated in Fig.~\ref{CohContour}. Below saturation there are large regions where the maximum achieved coherence goes beyond that of ultrastrong coupling, and in addition do not rely on resonance conditions. Beyond saturation however, these regions become extremely small and ultrastrong coupling is most efficient.

\begin{figure*}
   \centering
\includegraphics[width=2\columnwidth]{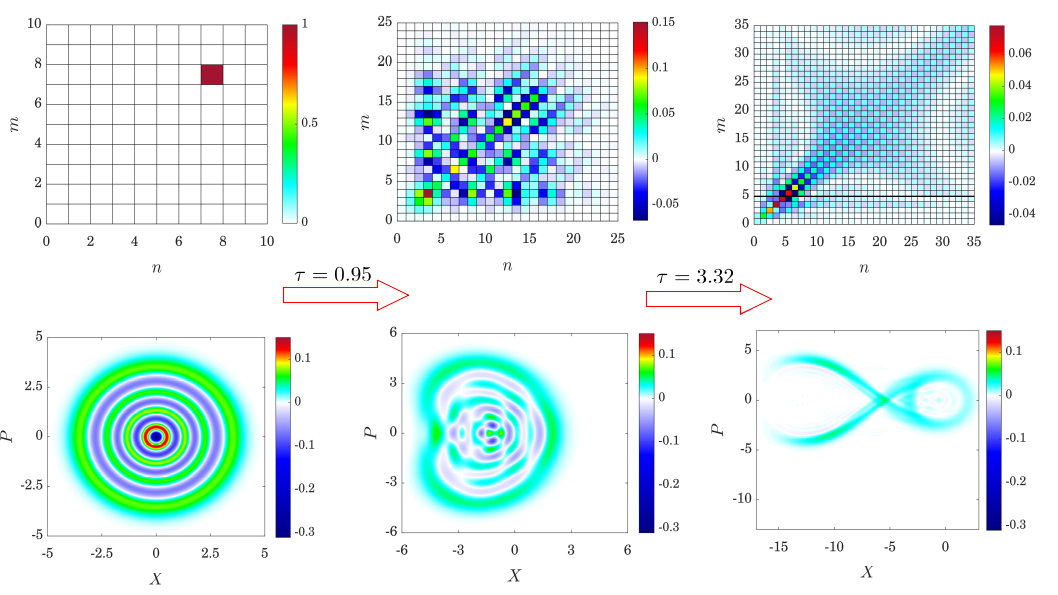}    \caption{A replica of Fig.~\ref{BigFig} from the main text with dephasing on the oscillator mode, at a rate 10\% that of $g^{(1)}$. The rightmost state corresponds to coherence, $\mathcal{C}=2.12$ while the central state corresponds to an example of a state with coherence $\mathcal{C}=1.21$.}
    \label{Decoh}
\end{figure*}

Decoherence naturally opposes the emergence of coherence from the combination of linear and nonlinear absorption. We illustrate this in Fig.~\ref{Decoh}. With dephasing on the oscillator at a rate 10\% that of $g^{(1)}$, quantum non-Gaussian coherence emerges, as seen in the negative regions of the central Wigner function, but is then suppressed by the decoherence. 

\begin{figure*}
   \centering
    \includegraphics[width=2\columnwidth]{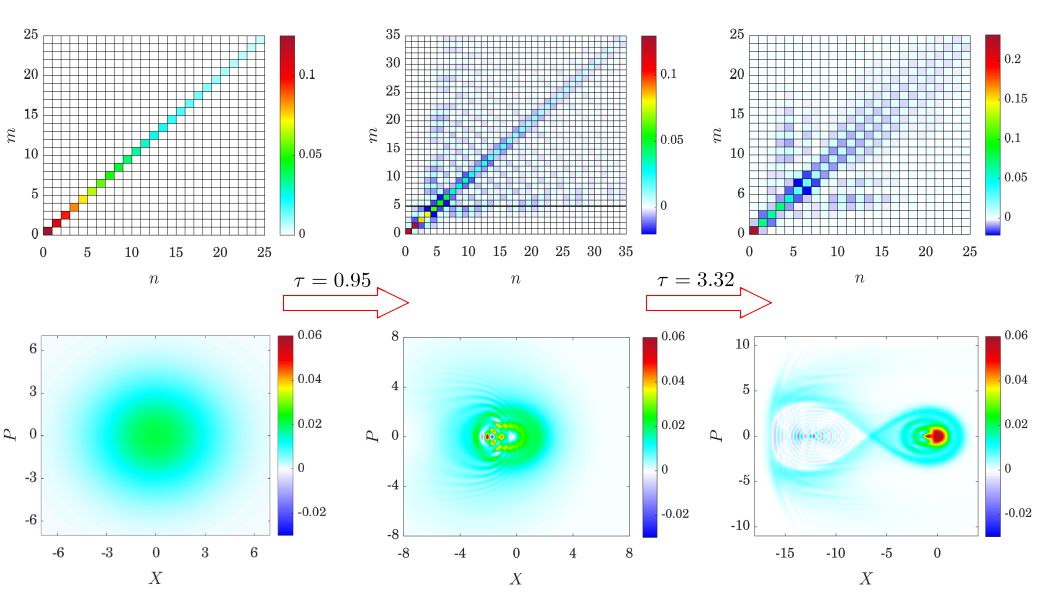}
    \caption{A replica of Fig.~\ref{BigFig} from the main text with a thermal state with mean occupation $\nb=7$.  The rightmost state corresponds to coherence, $\mathcal{C}=0.86$ while the central state corresponds to an example of a state with coherence $\mathcal{C}=0.27$.}
    \label{Therm}
\end{figure*}

\begin{figure*}
   \centering
    \includegraphics[width=2\columnwidth]{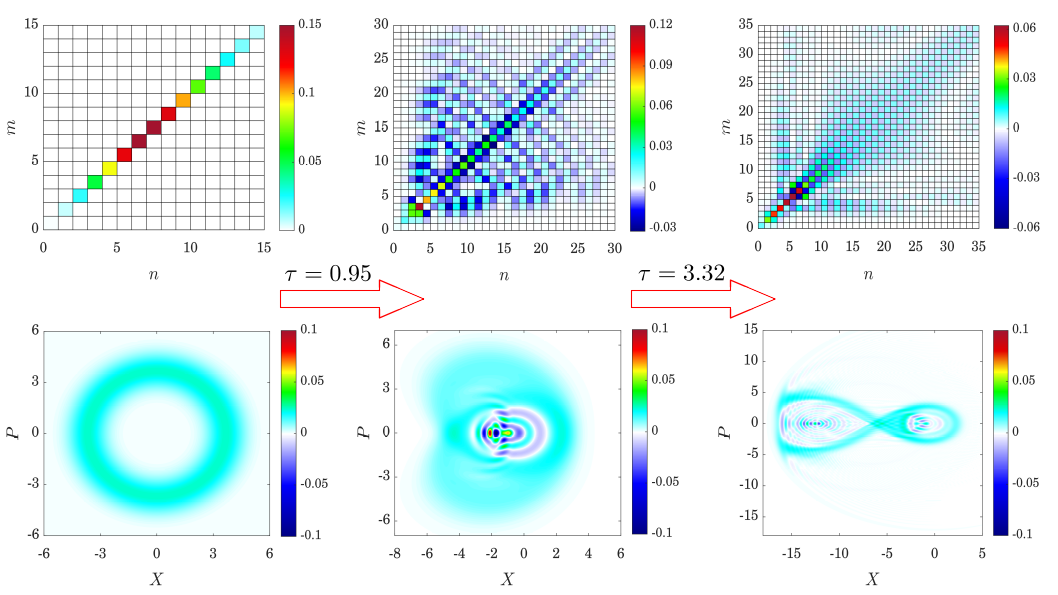}
    \caption{A replica of Fig.~\ref{BigFig} from the main text with a phase randomised coherent state with mean occupation $\nb=7$.  The rightmost state corresponds to coherence, $\mathcal{C}=1.9$ while the central state corresponds to an example of a state with coherence $\mathcal{C}=0.61$.}
    \label{PRCoh}
\end{figure*}

Replacing the highly nonclassical Fock states with diagonal states possessing thermal noise in the Fock basis does not eliminate the emergence of coherence, although as expected the amount of coherence and visibility of quantum non-Gaussian features reduces. We show thermal states with mean phonon number $\nb=7$ in Fig.~\ref{Therm}. Strikingly, despite having much lower coherence than the previous example generated under decoherence, the quantum non-Gaussian features are more robust to initial thermal noise in the Fock basis. These features are increased when the thermal state is replaced by phase randomised coherent states (Fig.~\ref{PRCoh}), which are also diagonal in the Fock basis but possess Poissonian noise.

\begin{figure*}
    \centering
\includegraphics[width=2\columnwidth]{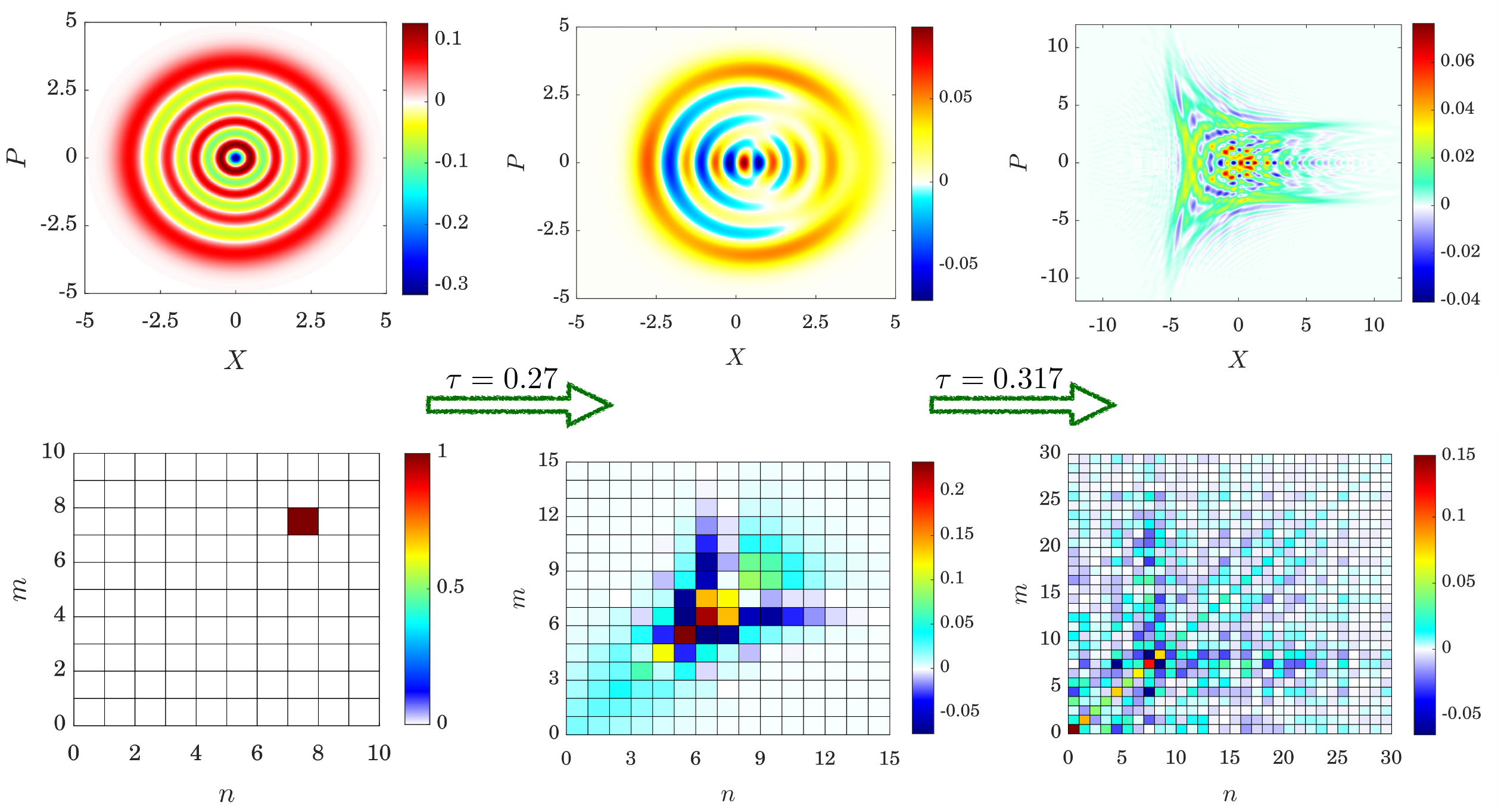}\\
    \caption{The emergence of quantum non-Gaussian coherence from frequency frustrated nonlinear absorption by an unsaturable absorber, driven by initially incoherent Fock states in $b$ at a fixed coupling ratio $\frac{g^{(2)}}{g^{(1)}}=0.1$. The initial ($\tau=0$) Fock state is $\ket{7}$, while the unsaturable absorber is initially in the ground state $\ket{0}$. The rightmost state corresponds to the maximum coherence, the central state corresponds to an example of a state with half the maximum coherence. The states are again radically non-Gaussian, with many negative regions and a lack of rotational symmetry. The corresponding density matrices show that coupling to an oscillator tends to generate states with superpositions clustered around the original Fock state with a large contribution from the ground state.}
    \label{BigFig2}
\end{figure*}

In an alternative model the absorber can be substituted by an oscillator mode $a$ with frequency $\Omega$, generating the Multi-Wave (MW) mixer interaction
\begin{equation}
    V^{(k)}_\text{MW}=g^{(k)}\left(a^\dagger b^k+a(b^\dagger)^k\right)\,.
\end{equation}
Again, these interactions preserve the total excitation number $N_\text{MW}=ka^\dagger a+b^\dagger b$ with a similar frequency condition to the qubit absorber system. The amount of coherence produced by their combination is significantly reduced compared to the qubit absorber, compared with similar parameters. However the Wigner functions, while qualitatively different, retain their complexity and quantum non-Gaussian features, again despite the impurity induced by the partial trace. The density matrices on the other hand show that quite different superpositions result. There is a much greater contribution from the classical ground state and fewer superpositions between Fock states with large number differences. Fig~\ref{BigFig2} shows a replica of the example states of Fig.~\ref{BigFig} for the MW interaction. 

\section{Coherence From Globally Active Transformations}

The frustration of the commutation relations means that the unitary evolution generated by the Hamiltonians $H_\omega+H_\Omega+V$ describes a globally active transformation in which the system gains energy from the interaction. In order for conservation of energy to hold, the interaction must be an approximation to a larger system from which this extra energy is drawn. Let us give a simple example of such a phenomenon. Conspicuously absent from our discussion of these features is the possibility of self-interactions of the oscillator which are active, such as single-mode squeezing. Squeezing changes the energy of the oscillator, and this energy must come from and go somewhere. In fact the squeezing self-interaction Hamiltonian is strictly an approximation to a larger two mode system which can actively mediate an exchange of energy via the downconversion of single photons into pairs. That is, the single-mode squeezer is a trilinear interaction in which the pump mode does not deplete and can be treated semi-classically~\cite{crouch_limitations_1988}. In fact let us dwell further on this example; the trilinear interaction alone does not produce coherence in the target oscillator (crucially, using our assumptions on preparation!), yet the semi-classical approximation of this interaction generates the single-mode squeezing interaction which does produce coherence. The trilinear interaction is globally passive, yet the single-mode squeezer is active. This is because the trilinear interaction represents an interaction depleting the external coherent drive, operating in a fully quantum mechanical way and is fully energy conserving among the real oscillator subsystems. In order to see the single-mode squeezer, one has to linearise the trilinear interaction by driving the pump mode strongly i.e. there is an approximately undepleted use of classical coherent states, which allows the process to gain energy and produce coherence.

\begin{figure}
    \centering
    \includegraphics[width=\columnwidth]{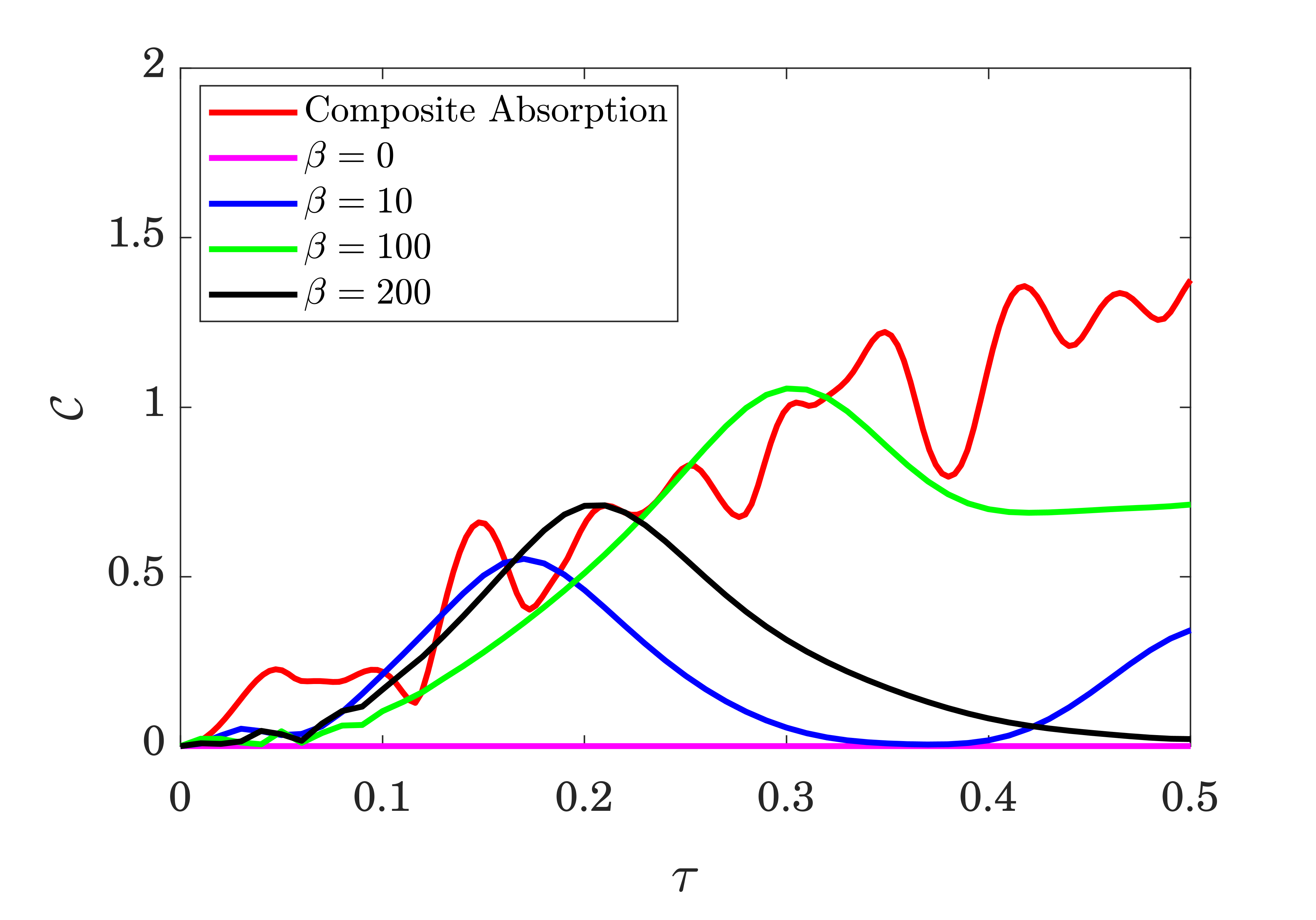}
    \caption{Coherence emerging from the completed model of the nonlinear absorption Hamiltonian, using the same parameters as previous examples. If $\beta=0$ there is no initial coherence and the passive interaction does not produce coherence in the oscillator. Strong classical pumping of mode $a$ allows coherence to emerge and approximates the nonlinear absorption for short times.}
    \label{modea}
\end{figure}

In the same way, the nonlinear absorption system can be seen as an incomplete description of a larger system. A candidate interaction Hamiltonian to complete the description involves the addition of an extra oscillator mode $a$, so that we have
\begin{equation}
    V_\text{compl}=g^{(1)}\left(\sigma_+ba+\sigma_-b^\dagger a^\dagger\right)+g^{(2)}\left(\sigma_+b^2+\sigma_-b^\dagger{}^2\right)
\end{equation}
Now the interaction involves a third mode with free Hamiltonian $H_\nu=\nu a^\dagger a$, and is again a passive interaction and does not produce local coherence from incoherent initial states. If instead the pump mode $a$ is driven by a coherent state $\ket{\beta}$, then the short-time coherence is retrieved. Fig.~\ref{modea} shows the oscillator coherence produced, taking the same parameters and initial states as before, but with the completionary mode $a$.

\section{Fock States in Admixture With Ground States}

\begin{figure*}
    \centering
    \includegraphics[width=2\columnwidth]{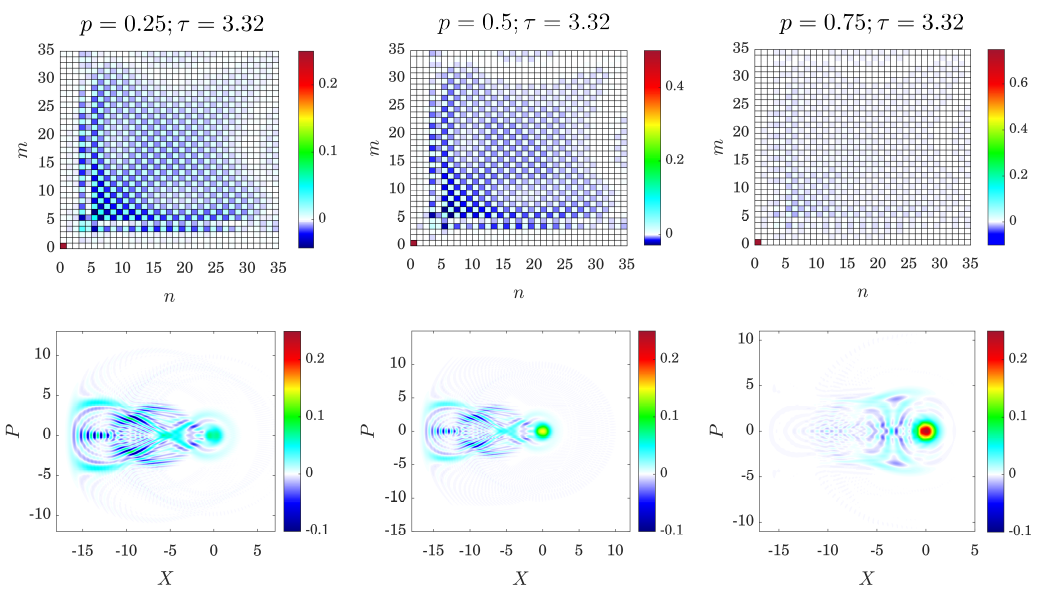}
    \caption{The maximum coherence when the initial state $\ket{7}$ is in an admixture with the ground state, for $p=0.25$, $p=0.5$ and $p=0.75$. This results in reduced coherence values $\mathcal{C}=3.02$, $\mathcal{C}=2.02$ and $\mathcal{C}=1$. The QNG features and spread into the Fock basis remain even for large contributions from the ground state.}
    \label{Atten}
\end{figure*}

The proposed methods to prepare Fock states from incoherent initial states and phase insensitive linear absorption~\cite{marek_deterministic_2016,slodicka_deterministic_2016} often result in an admixture of the ground state with the prepared Fock state $\ket{n}$, of the form
\begin{equation}
    p\ket{0}\bra{0}+(1-p)\ket{n}\bra{n}\,.
\end{equation}
This also does not prevent the quantum non-Gaussian features and coherence from emerging and also does not affect the timescale of their emergence. We present several examples in Fig.~\ref{Atten}, with $p=0.25$, $p=0.5$ and $p=0.75$. 

\end{document}